\documentclass[%
 aip,
 amsmath,amssymb,
 reprint,%
 groupedaddress,%
]{revtex4-1}
\usepackage{graphicx}
\usepackage{dcolumn}
\usepackage{bm}
\usepackage{cleveref}
\creflabelformat{equation}{#2#1#3}
\usepackage[utf8]{inputenc}
\usepackage[T1]{fontenc}
\usepackage{mathptmx}
\usepackage{etoolbox}
\usepackage{amsmath}
\usepackage{accents}
\newcommand{\utilde}[1]{%
  \ensuremath{\mathord{\vtop{\ialign{##\crcr
  $\displaystyle{#1}$\crcr
  \noalign{\kern1.5pt\nointerlineskip}
  $\sim$\crcr}}}}}
\begin{document}
\nocite{1,2,3,4,5,6,7,8,9,10,11,12,13,14,15,16,17,18,19,20,
21,22,23,24,25,26,27,28,29,30,31,32,33,34,35,36,37,38,39,40,
41,42,43,44,45,46,47,48,49,50,51,52,53,54,55,56,57,58,59,60,
61,62,63,64,65,66,67,68,69,70,71,72,73,74,75,76,77,78}
\preprint{AIP/123-QED}
\title[On the Special Theory of Relativity and Electromagnetism]{On the Special Theory of Relativity and Electromagnetism}
\author{O. L. de Lange}
\author{R. E. Raab}
\affiliation{Physics Department, University of Kwa-Zulu Natal, Pietermaritzburg, South Africa}
\date{\today}
\begin{abstract}
We present a formulation of the special theory of relativity which bears on F.A. Lindemann's assertion that this theory could have been reached “by pure logic soon after Isaac Newton”.  We start with the “intuitively plausible” pair of Galilean spatial transformations.  These simple relations possess a rich structure of ten properties.  From these, one discerns an axiomatic structure (and a synchrony convention) leading to the well-known Lorentz-type transformations which contain a universal constant, V\textsuperscript{2}.  Analysis of Fizeau's experiment (1851) shows that V\textsuperscript{2} = $c^2$, where $c$ is the speed of light \textit{in vacuo}.  Hence one obtains the Lorentz transformation.  Requisites for such a formulation (Galileo’s relativity principle, analytical mechanics, the method of changing a postulate, etc.) emerged during the 1600s and 1700s.  These observations provide a framework for Lindemann’s assertion.  We also consider inertially-moving systems of charge, and derive electromagnetic field equations and a force law by applying the Lorentz-type transformations to the theory of electrostatics.  The results are independent of any choice of units, and from their dependence on V\textsuperscript{2} one can infer how certain phenomena manifest in each of the three possible types of space-time. 

\end{abstract}

\maketitle

\section{\label{sec:intro}Introduction}

A striking aspect of the special theory of relativity is the large number of derivations that have been presented for the fundamental kinematic part of the theory (specifically, the Lorentz transformation equations) [\onlinecite{1,2,3,4}]. It is problematic to give an overall assessment of this situation: First, because subjective considerations may be involved [\onlinecite{1}].  And second, because --- as Berzi and Gorini noted (already in 1968) --- the “existing literature is very wide and rather unrelated and it would be almost impossible to give a fairly complete summary of it” [\onlinecite{2,5}].  The following brief remarks provide some essential background to our paper.

A common theme among many formulations of the special theory is a desire to simplify and clarify.  Einstein’s first (1905) paper on this theory  [\onlinecite{6}] was regarded as “notoriously difficult" [\onlinecite{7}], and in 1917 he published a “popular exposition” in which he “spared himself no pains in his endeavors to present the main ideas in the simplest and most intelligible form” [\onlinecite{8}]. From the trajectories $x = \pm ct$ and $x' = \pm ct'$ of light rays moving at a universal speed \textit{c} with respect to inertial frames in relative motion along common $xx'$-axes, the inferred linear relations between $(x',t')$ and $(x,t)$; and hence --- with the use of the relativity principle --- the Lorentz transformations.  A related approach, in which both the relativity principle and the light postulate are imposed on linear space and time transformations, has become standard [\onlinecite{9}].

Some authors found simplicity in a formulation based on the theory of electrodynamics.  Thus, according to Sommerfeld: “The path taken by Einstein in 1905... was steep and difficult.  The path we shall take is wide and effortless.  \textbf{It proceeds from the universal validity of the Maxwell equations}... It ends almost inadvertently at the Lorentz transformation and all of its relativistic consequences” [\onlinecite{10}].

Such formulations were criticized by Einstein. In 1952 he commented on a book by von Laue: “When one looks over your collection of proofs of the special relativity theory, one becomes of the opinion that Maxwell’s theory is unquestionable,” whereas, “in 1905 I already knew certainly that Maxwell’s theory” could not have generated validity because it did not account for the, “objective atomic structure of electromagnetic radiation” [\onlinecite{11}].  Elsewhere, Einstein emphasized that his theory of space and time has a more general validity than the theory of classical electrodynamics [\onlinecite{12,13}].

In Einstein’s theory, the fundamental role of a universal, limiting speed is played by the finite speed \textit{c} of light \textit{in vacuo} [\onlinecite{6}].  These properties of \textit{c} require a revision of the concept of time, and lead to a breakdown in ‘na\"ive visualizability’ [\onlinecite{14}] through the resulting relativity of simultaneity and time [\onlinecite{6}].  The reasoning can be challenging, as this excerpt from a well-known monograph illustrates: ``As to definitions of concepts we are, however, to some extent free and... it is possible to use such a definition of simultaneity that the velocity of light is constantly equal to $c$ in all inertial systems'' [\onlinecite{15}].  

It is therefore perhaps not surprising that attention should be given to single-postulate theories based on the relativity principle (‘relativity without light’).  What is surprising is the extent of this enterprise, which has continued to produce a large literature [\onlinecite{16,17}] since its inception in 1910 [\onlinecite{18,19}].

The intention (according to Lévy-Leblond) is to remove the historical link to a “restricted class of natural phenomena, namely, electromagnetic radiations”, so that special relativity becomes ``a sort of `super law' '' that constrains all the laws of physics [\onlinecite{20}].  Single-postulate derivations are independent of whether, strictly speaking, the photon mass is zero; and --- presumably --- of whether particles of zero mass actually exist [\onlinecite{20}].

These derivations yield so-called `V\textsuperscript{2} – Lorenz transformations' [\onlinecite{21}] (or ‘Lorentz-type transformations’) that contain an unknown universal constant V\textsuperscript{2} with dimensions  (velocity\textsuperscript{2}).  There are three distinct types of space and time transformations according to whether V\textsuperscript{2}  is > 0, < 0, or $\infty$ [\onlinecite{2,3,18,20}]. Thus one has the additional task of distinguishing between these possibilities on theoretical or experimental grounds [\onlinecite{2,3}].  In this connection, single-postulate theories have been criticized as being incomplete [\onlinecite{22,23}] 

With this background, we turn to the purpose of our paper.  Our motivation is succinctly conveyed by an opinion held by F.A. Lindemann (Lord Cherwell), who “often said that if only scientists had had their wits about them, they ought to have been able to reach the Relativity Theory by pure logic soon after Isaac Newton, and not to have to wait for the stimulus given to them by certain empirical observations that were inconsistent with the classical theory” [\onlinecite{24}].

To our knowledge, Lindemann did not provide any further details concerning this intriguing statement. Thus we consider the question: by what ‘pure logic’ could the special theory have been ‘reached’ during the 18th and 19th centuries?  In Section \ref{sec:kinematics} we present a simple kinematical analysis as a candidate for this purpose.  The resulting kinematics is that of the V\textsuperscript{2}– Lorentz transformations.  These transformations enable one to infer from the Fizeau experiment (1851) that to within experimental error $V^2 = c^2$, where \textit{c} is the speed of light \textit{in vacuo}.  Thus the Lorentz transformations are selected.  

In Section \ref{sec:historicalnotes} we consider some historical aspects that pertain to the question: is it reasonable to suppose that the kinematic theory of Section \ref{sec:kinematics} could have been formulated ‘soon after Newton?’

It is natural to wonder: if the special theory had preceded (rather than followed) the theory of classical electrodynamics, how might the development of the latter have been affected?  This is considered in Section \ref{sec:electrodynamics}, where we outline a derivation of the electromagnetic field equations (for inertially moving systems of charge), based on the V\textsuperscript{2} - Lorentz transformations and the law of electrostatic interaction (Coulomb’s law).  An expression for the force on a moving charge (the Lorentz force) is also derived.  The results are in forms that are independent of any specific choice of units --- thus their dependence on the universal constant V\textsuperscript{2} is associated only with the space and time transformations.  Consequently, one can readily ascertain how certain electromagnetic phenomena manifest in each of the three possible types of space-time. 

Our paper is intended to be didactic, and the various sections are written with this in mind.  We hope that the kinematic theory in Section \ref{sec:kinematics} will provide the reader with helpful insights on a remarkable and challenging part of physics.  We also hope that the notes in Section \ref{sec:historicalnotes} will encourage readers to consult the extensive literature on the historical aspects of this topic.   The analysis presented in Section \ref{sec:electrodynamics} could be a useful supplement to the standard formulations of electrodynamics.

\section{\label{sec:kinematics}Kinematics}

For simplicity, we work in one spatial dimension.  Thus we consider common coordinate axes $Ox$ and $O'x'$; the time along each is denoted by $t$ and $t'$, respectively, and $O'$ moves in the positive $x$-direction with constant speed $v$ relative to $O$.  At time $t = t' = 0$, the origins $O$ and $O'$ coincide.  We also suppose that a free (isolated) particle which is placed at rest on $O'x'$ (say) remains at rest there.  Such a particle moves with constant velocity along $Ox$.  Thus, if the law of inertia holds in one system, it holds in the other system (if $O'x'$ is inertial, so is $Ox$).  Our concern is with kinematics relative to these frames.  

In terms of elementary notions, $x' = x - d$ (where $d$ is the displacement of $O'$ relative to $O$) and $x = x' + d'$ (where $d'$ is the displacement of $O$ relative to $O'$).  For uniform relative motion, \\
\begin{equation}
\begin{split}
d = vt,    	\quad d' = vt'.	\label{eqn:1} \\		
\end{split}
\end{equation}

Hence one has the pair of Galilean transformations:
\begin{equation}
\begin{split}
x' &= x - vt,   			\label{eqn:2} \\	
\end{split}
\end{equation}
\begin{equation}
\begin{split}
x &= x' + vt'.     				\label{eqn:3} \\		
\end{split}
\end{equation}

These simple relations possess a surprisingly rich set of properties. Namely:
\begin{enumerate}
    \item The linearity of (\ref{eqn:2}) means that the change $dx'$ corresponding to changes $dx$ and $dt$ is independent of $x$ and $t$.  And similarly for $dx$ from (\ref{eqn:3}).  This is the homogeneity of inertial space: this space is ‘the same’, everywhere and at all times.
    \item Equations (\ref{eqn:2}) and (\ref{eqn:3}) are unaffected by the substitutions
\begin{equation}
\begin{split}
x \to -x, \quad x' \to -x', \quad v \to -v \, .		\label{eqn:4} \\	
\end{split}
\end{equation}
This is an instance of the isotropy of inertial space: for a given direction of the motion of $O'$ relative to $O$, it is immaterial which of the two directions along the coordinate axes one designates as positive.
    \item The dependence of $x'$ on $x$, $t$ in (\ref{eqn:2}) is the same as the dependence of $x$ on $x'$, $t'$ in (\ref{eqn:3}), apart from the necessary change in the sign of $v$.  This mathematical form invariance is an instance of the relativity principle (physical ‘laws’ have the same mathematical form in all inertial frames).
    \item According to (\ref{eqn:3}), the velocity of $O$ with respect to $O'$ is 
\begin{equation}
\begin{split}
v' = -v   .								\label{eqn:5} \\	
\end{split}
\end{equation}
This is the so-called reciprocity relation (or principle [\onlinecite{2}]).
    \item Addition of (\ref{eqn:2}) and (\ref{eqn:3}) gives
\begin{equation}
\begin{split}
t' = t   .   							\label{eqn:6} \\	
\end{split}             
\end{equation}
Thus time is the same, everywhere and in all inertial frames.  The transformations are also specified by (\ref{eqn:2}) and (\ref{eqn:6}), or by (\ref{eqn:3}) and (\ref{eqn:6}).
    \item The change $t \to -t$ (and hence $v \to -v$) leaves (\ref{eqn:2}) and (\ref{eqn:3}) unchanged.
    \item A fixed interval $\Delta x'$ along $O'x'$ has a value $\Delta x$ along $Ox$, determined by the positions of its endpoints on $Ox$ at the same instant $t$.  According to (\ref{eqn:2}),
\begin{equation}
\begin{split}
\Delta x' = \Delta x   .				    		\label{eqn:7} \\	
\end{split}
\end{equation}
That is, spatial intervals are frame-independent (invariant).
    \item And, from (\ref{eqn:6}), time intervals are invariant:
\begin{equation}
\begin{split}
\Delta t' = \Delta t.   							\label{eqn:8} \\	
\end{split}
\end{equation}
    \item The velocity of a particle along $Ox$ is u = $dx/dt$; along $O'x'$ it is $u' = dx' / dt'$.  From (\ref{eqn:3}) and (\ref{eqn:6}) we have:
\begin{equation}
\begin{split}
u = u' + v,						           	\label{eqn:9} \\	
\end{split}
\end{equation}

And the inverse relation,
\begin{equation}
\begin{split}
u' = u-v   .							           	\label{eqn:10} \\	
\end{split}
\end{equation}

The velocity-addition formula (\ref{eqn:9}) is a symmetric function of its two arguments; the inverse (\ref{eqn:10})  is an anti-symmetric function.
    \item The sequence of transformations from $Ox$ to $O'x'$ to $O''x''$ (which moves with velocity $u$ with respect to $O'x'$) is equivalent to a single transformation
\begin{equation}
\begin{split}
x'' = x - (v + u)t			,\qquad		t'' = t.			\label{eqn:11} \\	
\end{split}
\end{equation}
\end{enumerate}

We wish to find a path by which, starting from the transformations (\ref{eqn:2}) and (\ref{eqn:3}), we can reach the Lorentz transformations.  A portal is provided by the following question:
\begin{quote}
(I) What spatial transformations are allowed by the relativity principle?
\end{quote}
This question is well-posed if we require also properties 1 and 2 (the homogeneity and isotropy of inertial space), and the answer is an evident extension of (\ref{eqn:2}) and (\ref{eqn:3}):
\begin{equation}
\begin{split}
x' = \gamma (v)(x-vt)   ,							      	\label{eqn:12} \\	
\end{split}
\end{equation}
\begin{equation}
\begin{split}
x = \gamma (v)(x' + vt')   ,							    \label{eqn:13} \\	
\end{split}
\end{equation}
with $\gamma (-v) = \gamma (v)$.  Because $\gamma (v)$ is even and dimensionless, it must be a function of $v^2/V^2$, where $V$ is a constant with the dimensions of speed.  Also, $\gamma >0$ (because $x > vt$ in (\ref{eqn:12}) implies $x' > 0$), and $\gamma (0) = 1$.  Our task is to determine $\gamma (v)$.
The time transformations determined from (\ref{eqn:12}) and (\ref{eqn:13}) are
\begin{equation}
\begin{split}
t' = \gamma(v)\left[t - \frac{vx}{V^2(v)}\right],  \label{eqn:14} 
\\
\end{split}
\end{equation}
And its form-invariant inverse,
\begin{equation}
\begin{split}
t = \gamma(v)\left[t' + \frac{vx'}{V^2(v)}\right]  ,	\label{eqn:15} \\	
\end{split}
\end{equation}
where
\begin{equation}
\begin{split}
V^2(v) \equiv \frac{v^2 \gamma^2(v)}{\gamma^2(v) - 1}. \label{eqn:16}
\end{split}
\end{equation}
By inverting (\ref{eqn:16}) and taking the positive root, we have
\begin{equation}
\begin{split}
\gamma(v) = \frac{1}{\sqrt{1 - \dfrac{v^2}{V^2(v)}}}.	\label{eqn:17} \\	
\end{split}  
\end{equation}
Thus the question (I) has taken us to the Lorentz – type transformations (\ref{eqn:12}) and (\ref{eqn:14}), or (\ref{eqn:13}) and (\ref{eqn:15}) [\onlinecite{21}].  
To continue the analysis, we require the velocity-addition formula that follows from (\ref{eqn:13}) and (\ref{eqn:15}); and its inverse from (\ref{eqn:12}) and (\ref{eqn:14}).  Namely,
\begin{equation}                            
\begin{split}
u = \dfrac{u' + v}{1 + \dfrac{u'v}{V^2(v)}}, \label{eqn:18}				
\end{split}
\end{equation}
\begin{equation}
\begin{split}
u' = \dfrac{u - v}{1 - \dfrac{uv}{V^2(v)}}. \label{eqn:19}	
\end{split}
\end{equation}
Consider (\ref{eqn:19}), written (for brevity) as $u' = g(u,v)$.  Set $u = v_{PO}, \quad u' = v_{PO'}, \quad v = u_{O'O}$.  \text{ Then}
\begin{equation}
\begin{split}
v_{PO'} = g(v_{PO}, v_{O'O}) .\label{eqn:20}\\	
\end{split}
\end{equation}
Interchange the labels P and O', and use the reciprocity relation 
\begin{equation}
\begin{split}
v_{O'P} = -v_{PO'}. \label{eqn:21}\\	
\end{split}
\end{equation}
Then
\begin{equation}
\begin{split}
-v_{PO'} = g(v_{O'O}, v_{PO}). \label{eqn:22}\\	
\end{split}
\end{equation}

According to (\ref{eqn:20}) and (\ref{eqn:22}), $u' = g(u,v)$ is anti-symmetric in $u$ and $v$ [\onlinecite{25,26}].  (This important conclusion depends on the reciprocity relations (\ref{eqn:5}) and (\ref{eqn:21}).  We shall return to this shortly).  By inspection of (\ref{eqn:19}), this anti-symmetry requires
\begin{equation}
\begin{split}
V^2 (u) = V^2 (v)   .							                   \label{eqn:23} \\	
\end{split}
\end{equation}
Now $u$ and $v$ are arbitrary.  Therefore $V^2$ is a universal constant (the same for all inertial frames). 

Thus, from the question (I), and by the reasoning given, we have found three types of space and time transformations corresponding to $\gamma = 1$ (i.e., $V^2 = \infty$), $\gamma > 1$ (i.e., $V^2 > 0$), and $\gamma < 1$ (i.e., $V^2 < 0$).  These results were first obtained by von Ignatowsky (in 1910), who considered “at which... transformation equations one arrives when only the relativity principle is placed at the top of the investigation” [\onlinecite{18,19}].  Our intention here has been to provide an approach, based on the Galilean transformations (\ref{eqn:2}) and (\ref{eqn:3}), that could have led to the Lorentz-type transformations in the century or so after Newton.  We shall return to this aspect in what follows. 

The value of $V^2$ is determined from experiment.  We are interested in an early experiment that can be interpreted within our elementary kinematical framework.  A suitable candidate is the Fizeau experiment (1851).  Here a liquid flows with velocity $v$ through a tube, and the velocity $u$ of light in the liquid is measured relative to the laboratory-frame [\onlinecite{27}].  For parallel motion of light and liquid, the results are represented (to within experimental error) by 
\begin{equation}
\begin{split}
u &= \frac{c}{n} + v\left(1 - \frac{1}{n^2}\right), \label{eqn:24}
\end{split}
\end{equation}
where $n$ is the refractive index of the liquid. Now the addition formula (\ref{eqn:18}), with $u' = c/n$, gives
\begin{equation}
\begin{split}
u &= \frac{\dfrac{c}{n} + v}{1 + \dfrac{cv}{nV^2}}. \, \label{eqn:25} \\
\end{split}
\end{equation}
To compare (\ref{eqn:24}) and (\ref{eqn:25}), note that because $v \ll c$, (\ref{eqn:24}) can be written (to a good approximation) as
\begin{equation}
\begin{split}
u &= \dfrac{\dfrac{c}{n} + v}{1 + \dfrac{v}{nc}} \, \label{eqn:26}. \\
\end{split}
\end{equation}
According to (\ref{eqn:25}) and (\ref{eqn:26}),
\begin{equation}
\begin{split}
V^2 = c^2   .							            \label{eqn:27} \\	
\end{split}
\end{equation}
For experiments using anti-parallel light and liquid velocities, $v$ is replaced by $-v$ in (\ref{eqn:24}) --- (\ref{eqn:26}), and the same conclusion (\ref{eqn:27}) is reached.
Thus Fizeau’s experiment indicates that $V^2 > 0$; and it identifies (to within experimental error) the universal limiting speed $V$ in (\ref{eqn:12}), (\ref{eqn:14}), and (\ref{eqn:18}), with the speed $c$ of light \textit{in vacuo}.  With $V^2 = c^2$, (\ref{eqn:12}) and (\ref{eqn:14}) are the Lorentz transformations. 
The times $t$ and $t'$ are measured on clocks distributed along the $xx'$ – axes, and this raises a question regarding their synchronization.  We appeal again to the Galilean case.  Equation (\ref{eqn:2}) and, 
\begin{equation}
\begin{split}
x = x' - v't'   ,								                   \label{eqn:28} \\	
\end{split}
\end{equation}
Instead of (\ref{eqn:3}), also satisfy the relativity principle.  From (\ref{eqn:2}) and (\ref{eqn:28}), $v't' = -vt$, and thus the reciprocity relation (\ref{eqn:5}) implies frame-independent time (\ref{eqn:6}).
This suggests a synchronization procedure in which each freely-moving frame is used as a ‘signal’ to synchronize the clocks of the other frame.  Thus, clocks along $Ox$ are synchronized in such a way that the speed along $Ox$ of a fixed point on $O'x'$ is a constant $v$.  And conversely, clocks on $O'x'$ are synchronized so that the speed along $O'x'$ of a point on $Ox$ is also equal to $v$.  Thus, the reciprocity (\ref{eqn:5}) is imposed by means of a synchrony convention.  

This convention applies also in the derivation of the Lorentz-type transformations (\ref{eqn:12}) and (\ref{eqn:14}).  In fact, the quantities $x \mp vt$ in (\ref{eqn:2}), (\ref{eqn:3}), (\ref{eqn:12}), and (\ref{eqn:13}) could be regarded as ‘signatures’ of this convention.   The reciprocity (\ref{eqn:21}) requires also the synchronization of clocks in a frame in which $P$ is at rest.  
The foregoing synchrony was presented by Pars (1921), who emphasized:  “It is clear that we must introduce some convention in order to coordinate the two systems of measurement --- there would be no significance to the existence of an invariant velocity for observers whose systems of measurement were entirely independent” [\onlinecite{28}].

In this approach, the law of inertia is imposed by synchronizing clocks in such a way that the relative velocities of freely-moving objects are constant; and reciprocity relations are a convention [\onlinecite{29}].

The latter simplifies the theory, which would otherwise contain a parameter $v'/v$ - cf. (\ref{eqn:2}) and (\ref{eqn:28}).

Our main intention in this section is to demonstrate the extent to which a careful analysis of the ‘intuitive’ theory (\ref{eqn:2}), (\ref{eqn:3}) can inform and guide us to the decidedly non-intuitive Lorentz-type transformations (\ref{eqn:12}), (\ref{eqn:14}).  

Consider now the properties 1 --- 10. The Galilean relations (\ref{eqn:2}), (\ref{eqn:3}) follow from the three properties 1, 4, and 7: That is, spatial homogeneity, the reciprocity relation, and the frame-independence (\ref{eqn:7}) of spatial intervals. For the Lorentz-type formulations, the corresponding relations (\ref{eqn:12}) and (\ref{eqn:13}) follow from the last four properties 1 to 4: That is, spacial homogeneity, isotropy, form invariance (the relativity principle), and reciprocity.

If we refer to the frame-invariance of spatial intervals and the relativity principle as postulates, and to spatial homogeneity and isotropy as axioms, then the Galilean and Lorentz-type formulations are both 'single postulate' theories, resting on either one axiom (Galilean) or two axioms (Lorentz-type).
  
Note that the same synchrony convention (property 4) is used in both cases.  By contrast, the notion of an ‘infinitely rapid’ signal that synchronizes all clocks at a given instant (in a ‘flash’) is restricted to the Galilean case.  

Evidently, the foregoing simple theory of invariant spatial intervals introduces and elucidates various essential concepts, and provides useful clues to formulate and analyze the Lorentz-type transformations.  In essence, the restriction to absolute space is lifted, while maintaining the rest of the theory to the extent possible.  
The frame-dependence of spatial intervals when $\gamma \neq 1$ is deduced in the same way as in the Galilean case (property 7).  Thus from (\ref{eqn:12}), and by the same reasoning leading to (\ref{eqn:7}), we have
\begin{equation}
\begin{split}
\Delta x' = \gamma \Delta x   ,		                   			\label{eqn:29} \\	
\end{split}
\end{equation}
where $\Delta x'$ is an interval fixed on $O'x'$ and $\Delta x$ is its value along $Ox$.  Also, for time intervals, one has, from (\ref{eqn:14}) 
\begin{equation}
\begin{split}
\Delta t' = \gamma \Delta t   ,									\label{eqn:30} \\	
\end{split}
\end{equation}
where $\Delta t$ is a time interval on a clock fixed on $Ox$, and $\Delta t'$ the corresponding time difference measured on two separated clocks along $O'x'$.  Thus frame-dependence of spatial intervals requires frame-dependence of time intervals.
Note also that the reason for using the same functions $\gamma (v)$ in (\ref{eqn:12}) and (\ref{eqn:13}) is due to spatial isotropy and the reciprocity relation: from (\ref{eqn:29}) and its inverse it follows that a unit interval at rest on either $O'x'$ or $Ox$ must have the same value $\gamma^{-1}$ in the corresponding ‘stationary frame’ $Ox$ or $O'x'$, respectively.
This brings us to the question:  how reasonable is it to suppose that the Lorentz-type transformations (\ref{eqn:12}) and (\ref{eqn:14}) could have been reached by the preceding kinematic analysis (or, at least, parts of it) in the century or so after Newton?  Clearly, a certain level of mathematical analysis, and its application to analytical kinematics, are required here:  Cartesian axes, the notions of variable and function, and a dexterity for algebraic manipulation and interpretation.
A further aspect concerns motivation:  what could have prompted the early discovery of the Lorentz-type transformations?  As we mentioned, these were first obtained in 1910, as a response to the question (I) [\onlinecite{18,19}].  The motivation for (I) could be a desire to generalize, and it seems to us pointless to speculate on whether this should have arisen at an earlier stage.
Consider rather, instead of (I), the question:

\begin{quote}
(II) What transformations follow from (\ref{eqn:2}) and (\ref{eqn:3}) if we remove the restriction (property 7) to invariant spatial intervals?
\end{quote}

The intention here could be to establish an invariant property of space by assuming the contrary, and deducing consequences that are (seemingly) unacceptable or paradoxical --- such as a universal speed $V$ and the time dilation or contraction in (\ref{eqn:30}).  There is an interesting precedent here, because this technique was employed in work on physical geometry published in 1733 by Giralomo Saccheri.  We discuss these aspects in Section \ref{sec:historicalnotes}.

\section{\label{sec:historicalnotes}Historical Notes}

The publication of the three books of the \textit{Principia} in July 1687 marked the birth of Newtonian mechanics, and it was perhaps the high point of the Scientific Revolution in Europe.  Our concern is with matters related to our main theme; we consider first the influence of Newton, and then certain developments that occurred mostly after 1687.

\subsection{Newton}

We are interested in the style in which \textit{Principia} was written, and in Newton’s views on absolute space.  He acquired a proficiency in Cartesian (analytical) geometry as a student, and within a year or so (by 1666) became one of the foremost practitioners of the new analytics.  (“Never did seventeenth-century man build up so great a store of mathematical expertise, much of it of his own discovery, in so short a time” [\onlinecite{30}]).  The significance of coordinate geometry exceeded its utility in solving classical geometrical problems:  in fact, it “gave him his first true vision of the universalizing power of the algebraic free variable...” [\onlinecite{30}].

And yet, in 1684 he opted for the ‘geometry of the Ancients’: \textit{Principia} is written in the style of Euclid’s \textit{Elements} [\onlinecite{31}].  Evidently, Newton first studied classical geometry in detail only in 1680 (when he was almost forty).  The upshot was a vehement rejection of his Cartesian background, which he derided (for instance) as “the Analysis of the Bunglers in Mathematics”.  And he regretted his “mistake... in applying himself to the works of Des Cartes and other algebraic writers, before he had considered the elements of Euclide with that attention, which so excellent a writer deserves” [\onlinecite{31}].  (There was a twist to this some three decades later, when Newton claimed to have first proved most of the propositions  in \textit{Principia} via the new analytical geometry, and only subsequently reworking them, so “that the System of the Heavens might be founded upon good Geometry” [\onlinecite{32}].  There is, however, no documentary evidence to support this claim [\onlinecite{32}].)  We shall return to this matter of algebra versus geometry.

Newton’s views on absolute space were influenced by his predecessors.  Theories of space had a long history prior to Newton [\onlinecite{33,34}], and he incorporated this legacy “into his great synthesis and placed it as the concept of absolute space in the front line of physics” [\onlinecite{34}].  It is noteworthy that in \textit{Principia}, Newton managed to keep physics and metaphysics/religion “in separate compartments of his mind but for one exception, namely, his theory of space” [\onlinecite{35}].

Among the infinite number of spaces in which the laws of motion are valid, the “final degree of accuracy, the ultimate truth can be achieved only with respect to... absolute space” [\onlinecite{36}].  (Thus, instead of the fully reciprocal relations (\ref{eqn:2}) and (\ref{eqn:3}), one would stop at the ‘one-way’ relations 

\begin{equation}
\begin{split}
x' = x - x_A - vt_A, \qquad  t' = t_A,								\label{eqn:31} \\	
\end{split}
\end{equation}

where $x\textsubscript{A}$ is a coordinate in absolute space, and $t_\textsubscript{A}$ denotes Newton’s absolute time.) Newton identified absolute space as a space in which the center-of-gravity of the constituents of the solar system are at rest, as are the axes of Keplerian orbits.

Criticism of this doctrine of space came already from Huygens and Leibniz (two of Newton’s contemporaries), who asserted the equivalence of all inertial spaces or frames (as they were later termed).  It was recognized that dynamics in inertial space does not distinguish between these two points of view:  In modern terminology, the equation of motion is unaffected by (\ref{eqn:2}), (\ref{eqn:3}), and by (\ref{eqn:31}).  Disagreement arose concerning Newton’s claim that centrifugal force is a manifestation of rotation with respect to absolute space [\onlinecite{34}].

Huygens never wavered in his certainty that absolute space is a fiction.  He was so convinced that Newton was wrong on this score that he hoped it would be corrected in a second edition of \textit{Principia}.   (His views were expressed, \textit{inter alia}, in letters written in 1694 – the year before his death – to Leibniz.) [\onlinecite{34}].
Criticism of the idea of certain ‘absolutes’ came also from Berkeley, who questioned:  “Whether the notions of absolute time, absolute place, and absolute motion be not most abstractly metaphysical?  Whether it be possible for us to measure, compute, or know them?” [\onlinecite{37}].  And even Newton’s conviction was later queried: “We can indeed see from Newton’s formulation of it that the concept of absolute space made him feel uncomfortable...” [\onlinecite{38}].

By contrast, the relativity principle had an uneventful and successful passage, starting with its formulation (“with unrivalled clarity” [\onlinecite{39}]) by Galileo in the Dialogue concerning the two chief world systems (1632).  Newton’s statement of this principle, which he originally intended as a law of motion [\onlinecite{40}], appeared as a corollary in \textit{Principia}. 

Contrary opinions have been expressed on whether Galileo’s formulation of his principle is restricted to dynamics [\onlinecite{41}], or whether it applies to all phenomena.  Several authors favor the latter – among them Toretti [\onlinecite{39}], Brown [\onlinecite{42}], and Ohanian [\onlinecite{43}].

The explicit use of transformations between moving frames precedes \textit{Principia} by at least 20 years.  For instance, in his study of elastic impact (performed in 1667, and published posthumously in 1703), Huygens considered the collisions relative to two different frames --- a ship and the shore [\onlinecite{44}].

After \textit{Principia}, the science of mechanics was poised for further remarkable developments.  For these, the scene shifts to continental Europe.

\subsection{After Principia}

The 1700s and 1800s were a golden era for science in Europe.  In “France alone there were as many mathematicians of genius as Europe had produced in the preceding millennium” [\onlinecite{45}].  Much attention was paid to applications and reformulations of the new mechanics.  Our concern is with the latter; specifically, with the production of an analytical dynamics based on Descartes’ analytical geometry, Leibniz’s calculus, and the laws of motion.

The French savants “all worshiped ‘le grand Newton’, but they felt that his work needed to be remodeled... They proceeded to tear down what Newton had built, keeping only the foundations, and on these foundations they built a new, remodeled version of mechanics...” [\onlinecite{46}].  This process started already in the year that \textit{Principia} was published: in the decade to 1697, Pierre Varignon expressed parts of Book I of \textit{Principia} (dealing with central-force motion) in terms of analytical geometry and the calculus that Leibniz published in 1684 and 1686 [\onlinecite{47}].

Contributions from Hermann, the Bernoullis, and Euler followed.  The equation of motion was expressed in terms of differential equations, initially in two dimensions, with application to the attractive inverse-square force detailed by 1710 [\onlinecite{47}].

The culmination of the change from a geometric to an algebraic mechanics was marked by the publication of Lagrange’s \textit{Méchanique Analytique} (1788).  In his preference for algebra over geometry, Lagrange was the antithesis of the Newton of \textit{Principia}.  In the preface to his book, Lagrange wrote: “No diagrams will be found in this work.  The methods which I expound in it demand neither constructions nor geometrical or mechanical reasonings, but solely algebraic operations subjected to a uniform and regular procedure.  Those who like analysis will be pleased to see mechanics become a new branch of it, and will be obliged to me for having extended its domain” [\onlinecite{48}].

This development brought with it the enormous ‘economy of thought’ that algebra provides.  And it removed an impediment that was carried down from Greek mathematics, where all reasoning was given in words.  The extensive use of connected prose could “tax the patience of any reader... [familiar with] algebraic symbolism; and it was blamed for an ‘obscuring effect... in natural science’ that lasted into the 1800s [\onlinecite{49}].  At any rate, the kinematics in Section \ref{sec:kinematics} involves an essentially algebraic analysis: It would be at home in the \textit{Mecanique}, but not amongst the connected prose of \textit{Principia}.

With regard to absolute space, the criticisms by Huygens, Leibniz, and Berkeley played no evident role in the developments after \textit{Principia}.  In fact, Newton’s “essentially metaphysical conceptions of space, time... were then uncritically carried along with his scientific achievement and adopted by the European intellect” [\onlinecite{50}].

We turn now to a technique for generalizing a theory by altering one of the postulates on which it is based.  Our interest here is to draw an analogy between work on physical geometry (published by Giralomo Saccheri in 1733) and an aspect of the theory of space and time (the Lorentz-type transformations) discussed in Section \ref{sec:kinematics}.

Saccheri attempted to prove Euclid’s notorious fifth postulate (the postulate of parallels) [\onlinecite{51}].  Although the proof was flawed, his methodology has significance.  We require the following background.  Let \textit{P} be a point not on a given line \textit{L}.  And let \textit{N} be the number of lines, which can be drawn through \textit{P} (and lie in the plane of \textit{P} and \textit{L}), that do not intersect \textit{L}.  For the values of \textit{N} there are the following possibilities: (i) $N = 0$, (ii) $N \geq 2$, and (iii) the Euclidean case, $N = 1$.  Saccheri sought to obtain contradictions (with Euclid's other nine postulates) in the cases (i) and (ii); and hence infer the fifth postulate $(N = 1)$.  It turned out later that Saccheri’s results manifested instead certain aspects of non-Euclidean geometry [\onlinecite{52}]. 

In the theory of space and time in Section \ref{sec:kinematics}, it is the postulate of invariant (frame-independent) spatial intervals that is altered by introducing a positive, dimensionless factor $\gamma$ into (\ref{eqn:2}) and (\ref{eqn:3}): Cf. (\ref{eqn:12}) and (\ref{eqn:13}), and also (\ref{eqn:29}).  Then one has the following correspondence:  There are three possibilities for Saccheri’s number of parallels N, and also for the parameter $\gamma$ that specifies departure from invariant space.  And, for each of these possibilities there is a specific geometry or space-time, respectively.  Namely, Euclidean ($N = 1$), elliptic ($N = 0$), and hyperbolic ($N\geq 2$) geometries; and Galilean ($\gamma = 1$), Lorentzian ($\gamma > 1$), and ‘anti-Lorentzian' ($\gamma < 1$) space-times.  Also, both the non-Euclidean [\onlinecite{52}] and non-Galilean cases yield certain results that are ‘strange’ rather than contradictory.  (Here, ‘anti’ refers to the change from length contraction and time dilation, if $\gamma > 1$, to length dilation and time contraction if $\gamma < 1$; cf. (\ref{eqn:29}) and (\ref{eqn:30}).  And similarly for certain electromagnetic effects --- see Section \ref{sec:electrodynamics}).

To summarize: by the 1700s, an analytical mechanics had been developed (that is, an algebraic theory based on analytical geometry [\onlinecite{48}]).  Transformations between moving frames had been used in the solution of certain problems [\onlinecite{49}].  And the Galilean–Newtonian principle of relativity was known [\onlinecite{39, 42, 43}].

Thus, much of the framework necessary for the discovery of Lorentz-type transformations was available.  But two crucial pieces were missing:  The notion of equivalent inertial frames connected by the ‘Galilean’ transformations (\ref{eqn:2}) and (\ref{eqn:3}); and their modifications (\ref{eqn:12}) and (\ref{eqn:13}) to include relative spatial intervals (Section \ref{sec:kinematics}).  This raises the question:

\begin{quote}
(III) Were these steps within the capabilities of scientists in the century after Newton?
\end{quote}

To obtain (\ref{eqn:2}) and (\ref{eqn:3}) would require an ‘analytical mechanic’ with contrarian views on Newton’s absolute space.  We have mentioned the rapid development of analytical mechanics after \textit{Principia}; and Huygens and Leibniz were early dissenters on absolute space, who regarded the choice of inertial frame as essentially a matter of mathematical convenience [\onlinecite{51}].  Then, why did the transformations (\ref{eqn:2}) and (\ref{eqn:3}) appear in physics only much later? [\onlinecite{54}] Perhaps it was the distraction of so many competing, and evidently more promising, attractions (applications and alternative formulations of mechanics)?

Jammer emphasizes: “how little the actual progress of mechanics was influenced by general considerations concerning the nature of absolute space.  Among the great French writers on mechanics... none of them was much interested in the problem of absolute space.  They all accepted the idea of a working hypothesis without worrying about its theoretical justification.  In reading... their works, one discovers that they felt that science could very well dispense with general considerations about absolute space” [\onlinecite{55}].  (The change – when it eventually came – was dramatic and succinctly expressed:  In the Introduction to his 1905 paper, Einstein rejected the notion of a space “endowed with special properties” as “superfluous’'[\onlinecite{6}].)

For the second step – from (\ref{eqn:2}), (\ref{eqn:3}) to (\ref{eqn:12}) and (\ref{eqn:13}) - a ‘Saccheri of space and time’ could have sufficed.  The method of changing a postulate was known since at least 1733.  But there is no indication that the relevant question (II) was ever considered at all.

On the other hand, by the 1800s, there were proposals concerning even more intricate properties of space, and a possible role in dynamics.  The first concerns Riemann’s analysis of the ‘local’ behavior of space.  Rindler reminds us that “Riemann, in his celebrated inaugural lecture of 1854, had already suggested that the differential geometry of our three-space might be determined by ‘external forces’”.  And he adds: “It is fascinating to speculate on the course that physics might have taken if Riemann had lit upon gravity as the curver of space.  A geodesic law of motion would not have been too far-fetched even then” [\onlinecite{56}].

Even more remarkable was the suggestion by Clifford in 1876 of a spatio-temporal variation of such curvature caused by matter and its motion [\onlinecite{57}].  His work anticipated the union of geometry and physics, and it “is hard to realize that it was published forty years before Einstein announced his theory of gravitation” [\onlinecite{58}].  In this general theory of relativity, the curvature of space-time is related to the distribution of gravitating mass and energy.  Far from massive objects, in regions of low mass-density, the pseudo-Euclidean space-time of special relativity is a good approximation.

Of his two creations in relativity, Einstein remarked: “Compared with this problem [general relativity], the original [special] relativity is child’s play” [\onlinecite{59}].  While Riemann and Clifford speculated at the fringes of general relativity, this ‘child’s play’ continued to go unnoticed.

Based on the preceding historical notes, it seems reasonable to conclude that something like the kinematic theory of Section \ref{sec:kinematics} could have been formulated in the century or so after Newton (cf. a question posed in Section \ref{sec:intro}).  That is, Lindemann’s assertion regarding this matter seems well-founded.

Evidently, the questions (I) and (II) are instances where “posing a problem is a much finer art than its solution” [\onlinecite{60}].  As to why Newton’s successors failed in this, the situation is unclear (and perhaps ‘unknowable’).

\section{\label{sec:electrodynamics}Electrodynamics}

It is natural to wonder how the development of physics might have been affected if the discovery of Lorentz-type transformations had preceded that of the microscopic electromagnetic field equations. We consider one aspect of this topic here by outlining a derivation of the field equations for inertially-moving systems of charge.

The Lorentz transformations were discovered (around the turn of the 19th century) in studies of electromagnetism, and the special theory of relativity was formulated soon after [\onlinecite{12,27}]. Later, Rosser [\onlinecite{61}] considered the opposite procedure (of `electromagnetism via relativity'): he derived the field equations (in SI units, and for inertially-moving systems) by applying the Lorentz transformations to the theory of electrostatics. An expression for the Lorentz force on a moving charge was also deduced.

We are concerned with the theory at an (imagined) earlier stage, where the space and time transformations depend on an unknown universal constant $V^2$ (Section II), and systems of electromagnetic units had not yet been devised. Then the field equations and Lorentz force (obtained by applying the Lorentz-type transformations) have two helpful features:

\begin{enumerate}
\item They are independent of any particular choice of a system of units.
\item Their dependence on $V^2$ is due entirely to the theory of space and time.
\end{enumerate}

(See (\ref{eqn:40})--(\ref{eqn:44}).) Consequently, from these equations one can infer how certain electromagnetic phenomena manifest in each of the three possible types of space-time.

The benefit of a `unit-free' formulation was emphasized by Heras in a study of the Galilean limits of electrodynamics [\onlinecite{62}]. For example, to exclude situations where a physical result depends on a choice of units. The details can be subtle, but the essential point is that a choice of units may alter the dependence of electrodynamic equations on $V^2$.

We start our derivations with a few remarks on the theory of electrostatics. This theory is contained in the two field equations,
\begin{equation}
\nabla \cdot \mathbf{E}(\mathbf{r}) = \alpha \rho(\mathbf{r}) \, , \label{eqn:32}
\end{equation}
\begin{equation}
\nabla \times \mathbf{E}(\mathbf{r}) = 0 \, , \label{eqn:33}
\end{equation}
plus appropriate boundary conditions. Here $\alpha$ is a constant, and $\mathbf{E}(\mathbf{r})$ is the electric field of a static electric charge density $\rho(\mathbf{r})$.

These field equations follow from an experimental law (Coulomb's Law, 1785) for the electrostatic interaction between charges, and the definition of electric field in terms of the force
\begin{equation}
\mathbf{F} = q\mathbf{E} \label{eqn:34}
\end{equation}
on a point charge $q$. This law and definition show that the field of a point charge $Q$ located at the origin of coordinates can be written
\begin{equation}
\mathbf{E} = \frac{\alpha Q}{4\pi r^2} \hat{\mathbf{r}} \, , \label{eqn:35}
\end{equation}
where $r^2 = x^2 + y^2 + z^2$, $\hat{\mathbf{r}} = \mathbf{r}/r$, and the factor $4\pi$ is included for convenience.

For the total field of a system of charges lying within a closed surface $S$, it can be shown that [\onlinecite{63}]
\begin{equation}
\oint_S \mathbf{E} \cdot d\mathbf{S} = \alpha Q_T \, , \label{eqn:36}
\end{equation}
where $Q_T$ is the total charge within $S$. (This result depends on the inverse-square, central properties evident in (\ref{eqn:35}).) The differential form corresponding to (\ref{eqn:36}) is (\ref{eqn:32}). And (\ref{eqn:33}) is a consequence of the conservative property of (\ref{eqn:35}) [\onlinecite{63}].

This theory applies in the rest-frame $K'$ of the system of charge. Let $K$ be a frame in which the charges move with the same constant velocity $\mathbf{v}$. We seek the field equations and the force equation in $K$.

First note that for the charge density $\rho(\mathbf{r},t)$ and for the electric field $\mathbf{E}(\mathbf{r},t)$ in $K$, the space and time derivatives are related by [\onlinecite{64}]
\begin{equation}
\frac{\partial}{\partial t} = -\mathbf{v} \cdot \nabla \, . \label{eqn:37}
\end{equation}

For the charge density, (\ref{eqn:37}) yields
\begin{equation}
\frac{\partial \rho}{\partial t} + \nabla \cdot \mathbf{J} = 0 \, , \label{eqn:38}
\end{equation}
where
\begin{equation}
\mathbf{J} = \rho \mathbf{v} \label{eqn:39}
\end{equation}
is the electric current density. Thus the continuity equation (\ref{eqn:38}) is automatically satisfied in inertially-moving systems.

The field equations in $K$ depend on $\mathbf{E}(\mathbf{r},t)$, $\rho(\mathbf{r},t)$, and the constants $\mathbf{v}$ and $V^2$. We first write down these equations, and the force equation, and then outline how they are obtained.

\begin{equation}
\nabla \cdot \mathbf{E} = \alpha \rho \, , \label{eqn:40}
\end{equation}
\begin{equation}
\nabla \times \mathbf{E} = -\frac{1}{V^2} \frac{\partial}{\partial t} \mathbf{v} \times \mathbf{E} \, , \label{eqn:41}
\end{equation}
\begin{equation}
\nabla \cdot (\mathbf{v} \times \mathbf{E}) = 0 \, , \label{eqn:42}
\end{equation}
\begin{equation}
\nabla \times (\mathbf{v} \times \mathbf{E}) = \alpha \mathbf{J} + \frac{\partial \mathbf{E}}{\partial t} \, . \label{eqn:43}
\end{equation}

And the force on a point charge $q$ that moves with velocity $\mathbf{u}$ in $K$ is
\begin{equation}
\mathbf{F} = q\mathbf{E} + \frac{1}{V^2} q\mathbf{u} \times (\mathbf{v} \times \mathbf{E}) \, . \label{eqn:44}
\end{equation}

The task of deducing the field equations is shortened by noting that (\ref{eqn:42}) follows from (\ref{eqn:41}) with the use of an identity:
\begin{equation}
\begin{split}
\nabla \cdot (\mathbf{v} \times \mathbf{E}) &= -\mathbf{v} \cdot (\nabla \times \mathbf{E}) \\
&= \frac{1}{V^2} \frac{\partial}{\partial t} \mathbf{v} \cdot (\mathbf{v} \times \mathbf{E}) \\
&= 0 \, .
\end{split} \label{eqn:45}
\end{equation}
(The first step is an identity, the second step follows from (\ref{eqn:41}) and the constancy of $\mathbf{v}$, and the third from orthogonality of $\mathbf{v}$ and $\mathbf{v} \times \mathbf{E}$.) And similarly, from the identity
\begin{equation}
\nabla \times (\mathbf{v} \times \mathbf{E}) = \mathbf{v}(\nabla \cdot \mathbf{E}) - (\mathbf{v} \cdot \nabla)\mathbf{E} \, , \label{eqn:46}
\end{equation}
and (\ref{eqn:37}), (\ref{eqn:40}), and (\ref{eqn:39}), we obtain (\ref{eqn:43}).

We prove (\ref{eqn:40}) by establishing the frame-independence of the infinitesimals $\mathbf{E} \cdot d\mathbf{S}$ in (\ref{eqn:36}) [\onlinecite{65}]. Here we require the invariance of the coordinates perpendicular to the direction of motion along the $x$-axis [\onlinecite{66}]:
\begin{equation}
y' = y \, , \quad z' = z \, . \label{eqn:47}
\end{equation}

From (\ref{eqn:29}) and (\ref{eqn:47}), the components of the surface elements $d\mathbf{S}$ (that are parallel and perpendicular to $\mathbf{v}$) transform according to
\begin{equation}
dS_{\parallel} = \frac{1}{\gamma} dS_{\parallel}' \, , \quad dS_{\perp} = dS_{\perp}' \, . \label{eqn:48}
\end{equation}

Also, the force $\mathbf{F}'$ in $K'$ and the force $\mathbf{F}$ on a charge $q$ that is instantaneously at rest in $K$ are related by (see Appendix)
\begin{equation}
\mathbf{F} = (F_x', \, \gamma F_y', \, \gamma F_z') \, , \label{eqn:49}
\end{equation}
and therefore so are $\mathbf{E}$ and $\mathbf{E}'$ (the forces per unit charge). That is,
\begin{equation}
E_{\parallel} = E_{\parallel}' \, , \quad E_{\perp} = \gamma E_{\perp}' \, . \label{eqn:50}
\end{equation}

From (\ref{eqn:48}) and (\ref{eqn:50}),
\begin{equation}
E_{\perp} \, dS_{\parallel} = E_{\perp}' \, dS_{\parallel}' \, , \quad E_{\parallel} \, dS_{\perp} = E_{\parallel}' \, dS_{\perp}' \, . \label{eqn:51}
\end{equation}
Thus (\ref{eqn:36}) applies also to the electric field $\mathbf{E}(\mathbf{r},t)$ of inertially-moving charge. The corresponding differential form is (\ref{eqn:40}).

To prove the remaining field equation (\ref{eqn:41}), we require an explicit expression for the electric field $\mathbf{E}$ of an inertially-moving point charge $Q$. We start with the electrostatic field of $Q$ when located at the origin $O'$ of its rest frame $K'$, cf.\ (\ref{eqn:35}):
\begin{equation}
\mathbf{E}' = \frac{\alpha Q}{4\pi (x'^2+y'^2+z'^2)^{3/2}} (x',y',z') \, . \label{eqn:52}
\end{equation}

From $\mathbf{E}'$ we obtain $\mathbf{E}$ in two steps. First, we convert to the coordinates of $K$, using $x' = \gamma x$, $y' = y$, $z' = z$. Second, we include a factor $\gamma$ in the $y$- and $z$-components of (\ref{eqn:52}); cf.\ (\ref{eqn:49}). Then
\begin{equation}
\mathbf{E} = \frac{\alpha Q \gamma}{4\pi(\gamma^2 x^2 + y^2 + z^2)^{3/2}} (x,y,z) \, . \label{eqn:53}
\end{equation}

A short calculation (in Cartesian coordinates) shows that
\begin{equation}
\nabla \times \mathbf{E} = 3(\gamma^2 - 1) \frac{\alpha Q \gamma}{4\pi(\gamma^2 x^2 + y^2 + z^2)^{5/2}} (0, xz, -xy) \, . \label{eqn:54}
\end{equation}

Also, with $\mathbf{v} = (v,0,0)$, (\ref{eqn:37}) and (\ref{eqn:53}) give
\begin{equation}
\begin{split}
\frac{\partial}{\partial t} \mathbf{v} \times \mathbf{E} &= -v \frac{\partial}{\partial x} \frac{\alpha Q \gamma}{4\pi(\gamma^2 x^2+y^2+z^2)^{3/2}}(0,-vz,vy) \\
&= -3v^2\gamma^2 \frac{\alpha Q \gamma}{4\pi(\gamma^2x^2+y^2+z^2)^{5/2}}(0,xz,-xy) \, . \label{eqn:55}
\end{split}
\end{equation}

According to (\ref{eqn:16}), $v^2\gamma^2 = V^2(\gamma^2-1)$. Hence (\ref{eqn:54}) and (\ref{eqn:55}) yield (\ref{eqn:41}). This derivation is readily extended to a system of charges [\onlinecite{67}].

In the derivation of the force equation (\ref{eqn:44}), it is helpful to start with the special case of parallel motion of $q$ and $Q$. That is, $\mathbf{u} = (u_x,0,0)$. Then (see Appendix)
\begin{equation}
\mathbf{F} = (F_x', \, \gamma(1-u_xv/V^2)F_y', \, \gamma(1-u_xv/V^2)F_z') \, , \label{eqn:56}
\end{equation}
where $\mathbf{F}' = q\mathbf{E}'$. By the same two steps that lead from (\ref{eqn:52}) to (\ref{eqn:53}), we can express (\ref{eqn:56}) as
\begin{equation}
\mathbf{F} = q\mathbf{E} + \frac{1}{V^2} q(0, \, -u_xvE_y, \, -u_xvE_z) \, , \label{eqn:57}
\end{equation}
where $\mathbf{E}$ is the electric field (\ref{eqn:53}) in $K$. The vector in parenthesis is $u_x(\mathbf{v} \times \mathbf{E})$ when $\mathbf{u} = (u_x,0,0)$ and $\mathbf{v} = (v,0,0)$.

In general, $\mathbf{u} = (u_x,u_y,u_z)$ and the vector in parenthesis in (\ref{eqn:57}) is replaced by (see Appendix)
\begin{equation}
(u_yvE_y + u_zvE_z, \, -u_xvE_y, \, -u_xvE_z) \, , \label{eqn:58}
\end{equation}
which is $\mathbf{u} \times (\mathbf{v} \times \mathbf{E})$ when $\mathbf{v} = (v,0,0)$. Thus we have obtained (\ref{eqn:44}).

With the magnetic field defined by
\begin{equation}
\mathbf{B} = \frac{1}{k} \mathbf{v} \times \mathbf{E} \label{eqn:59}
\end{equation}
(where $k$ is a constant), (\ref{eqn:40})--(\ref{eqn:44}) are the field and force equations of microscopic electromagnetism: they are independent of any particular choice of units, and (as it turns out) they apply also to accelerated charge [\onlinecite{68}].

We are interested in using this theory to discriminate (in the first instance) between the types of space-time corresponding to $V^2$ positive, negative, or infinite. Thus we consider the magnetic force $q\mathbf{u} \times (\mathbf{v} \times \mathbf{E})/V^2$ in (\ref{eqn:44}). Here $\mathbf{E}$, given by (\ref{eqn:53}), is along the position vector of $q$ with respect to $Q$. It follows that for $V^2 > 0$ parallel electric currents attract, and anti-parallel currents repel. If $V^2 < 0$ then the opposite is true. Experiment readily confirms the former: the finiteness of $V^2$, and its positive sign, can be `sensed with one's hands.' (Experimental studies of the interaction between current-carrying elements were performed by Amp\`ere between 1820 and 1826 in his monumental work that laid the foundations of electrodynamics [\onlinecite{69}].)

It is useful to express two standard results in terms of our unit-free formulation, namely, Amp\`ere's Law
\begin{equation}
\oint_C \mathbf{B} \cdot d\mathbf{l} = (\alpha/k) I \, , \label{eqn:60}
\end{equation}
and the magnetic force
\begin{equation}
d\mathbf{F} = (k/V^2) I \, d\mathbf{l} \times \mathbf{B} \label{eqn:61}
\end{equation}
on an infinitesimal current-carrying element $d\mathbf{l}$. These are obtained from (\ref{eqn:43}), (\ref{eqn:59}) and (\ref{eqn:44}), (\ref{eqn:59}), respectively. It follows from (\ref{eqn:60}) and (\ref{eqn:61}) that the attractive force between two long parallel currents $I$ is
\begin{equation}
F = \frac{\alpha}{V^2} \frac{I^2 L}{2\pi d} \, , \label{eqn:62}
\end{equation}
where $L$ is the length of each current, and $d$ ($\ll L$) is their separation.

The determination of $V$ is concerned with the constant $\alpha/V^2$ in (\ref{eqn:62}), and with the constant $\alpha$ in the electrostatic force
\begin{equation}
F = \alpha \frac{Q^2}{4\pi r^2} \label{eqn:63}
\end{equation}
between point charges $Q$. The procedure --- which involves two tasks --- is as follows [\onlinecite{62}]. First, a unit of charge is defined by specifying either $\alpha$ or $\alpha/V^2$. Then the value of the other constant is determined by experiment, and a value for $V$ is obtained.

For instance, in SI units the unit of electric current $I$ (the ampere A) --- and hence of charge --- is defined by assigning to the permeability of free space $\mu_0$ (our $\alpha/V^2$) the value $4\pi \times 10^{-7} \, \text{N} \, \text{A}^{-2}$. The permittivity of free space $\epsilon_0$ ($=1/\alpha$) can be determined from capacitance measurements [\onlinecite{70}]. The resulting value of $V$ ($= 1/\sqrt{\epsilon_0 \mu_0}$) is equal (to within experimental error) to the speed $c$ of light \textit{in vacuo}. In Gaussian and Heaviside-Lorentz units it is the other way round: values of $\alpha$ are assigned (equal to $4\pi$ and $1$, respectively) and measurements of $\alpha/V^2$ yield again $V = c$. Thus (to within experimental error, and as required of a universal constant), $V$ is independent of which of these three definitions of charge one adopts [\onlinecite{62}].

In fact, it is possible (at least in principle) to measure $V$ using electric and magnetic forces, but without defining a unit of electric charge at all. Consider a device in which the electrostatic attraction between the plates of a charged parallel-plate capacitor,
\begin{equation}
F_e = \alpha \frac{Q^2}{2A} \, , \label{eqn:64}
\end{equation}
is balanced by the repulsive magnetostatic force between two overlapping square arrays of anti-parallel current-carrying wires,
\begin{equation}
F_m = \frac{\alpha}{V^2} \frac{1}{2} N^2 I^2 \, . \label{eqn:65}
\end{equation}
Here $A$ ($=L^2$) is the area of a plate, $N$ is the number of wires in an array, and $I$ is the current in a wire. (These forces are deduced from (\ref{eqn:34}), (\ref{eqn:36}) and (\ref{eqn:60}), (\ref{eqn:61}), respectively.)

Equality of $F_e$ and $F_m$ gives $V = NLI/Q$. The ratio $I/Q$ can be determined by two electrolysis experiments: $I/Q = m_I/Tm_Q$, where $m_I$ is the mass deposited in a time $T$ by the current $I$, and $m_Q$ is the mass deposited by discharging the capacitor into the electrolyte. Thus
\begin{equation}
V = N \frac{m_I}{m_Q} \frac{L}{T} \label{eqn:66}
\end{equation}
is determined from a length divided by a time, and a ratio of two masses. (Typically, $m_Q \ll m_I$ here. But, if the capacitor is repeatedly charged and discharged at a frequency $\nu$, and the resulting current $\nu Q$ flows for a time $T$, then the mass deposited is enhanced by a factor $\nu T$.)

$V^2$ appears also in (\ref{eqn:41}), and if $V^2 = \infty$ then $\nabla \times \mathbf{E} = 0$. And, in source-free regions (where $\rho = 0$), (\ref{eqn:40})--(\ref{eqn:43}) and (\ref{eqn:59}) yield the differential equation
\begin{equation}
\left(\nabla^2 - \frac{1}{V^2}\frac{\partial^2}{\partial t^2}\right)(\mathbf{E} \text{ or } \mathbf{B}) = 0 \, , \label{eqn:67}
\end{equation}
which has travelling-wave solutions if $V^2 > 0$, but not if $V^2 < 0$ or $V^2 = \infty$.

A feature in some of the foregoing analysis is the intricacy of the inputs versus the simplicity of the outputs. For instance, the interaction (\ref{eqn:62}) is obtained from force transformations that depend on all three aspects of the Lorentz-type transformations (relative space, time, and simultaneity), and the work-energy theorem for the energy $\gamma m V^2$ (see Appendix). And the synchronization convention introduced in the kinematics (Section II) is implicit here: for example, in the field equations (\ref{eqn:40})--(\ref{eqn:43}), which involve simultaneous values of field components at different spatial points.

In Einstein's theory, clocks are synchronized in such a manner that the `one-way' speeds of light \textit{in vacuo} (and hence also the `two-way' speed, $c$) are equal [\onlinecite{6}]. And $c$ (which can be measured using a single clock) is identified as a universal constant.

There are many ways to measure $c$ [\onlinecite{71}]. In 1898, Poincar\'e wrote that the ``postulate'' of isotropic light velocity ``could be contradicted by [experience]\ldots if results of diverse measurements were not harmonious.'' And he added: ``We ought to regard ourselves fortunate that this contradiction does not take place\ldots'' [\onlinecite{72}].

\section{\label{sec:Discussion}Discussion}
In later years, Einstein became convinced, "that pure mathematical construction enables us to discover the concepts [of Physics] and the laws connecting them..." [\onlinecite{73}]. In an interesting commentary, Pais suggested that Einstein's stance on mathematical formalism may have originated with the 1905 paper [\onlinecite{6}], whose, "kinematic part... has the ideal axiomatic structure of a finished theory, a structure which had abruptly dawned on him [Einstein] after a discussion with [his friend] Besso" [\onlinecite{74}]. And Pais continued: "Is it possible that this experience was so overwhelming that it seared his mind and partially blotted out reflections and information that had been with him earlier, as the result of deep-seated desires to come closer to the divine form of pure creation?"  These sentiments convey, in an evocative manner, part of the motivation for our work --- on a notion that the special relativity theory should have been found by pure logic long before 1905 (see Lindemann's remark in Section I).

To this end, we have presented a formulation of special relativity based on reasoning that could have been used (at various stages) during the 18th and 19th centuries (Section II and III).  The analysis is essentially algebraic and therefore belongs to the period after publication of Lagrange’s \textit{Mecanique Analytique} in 1788.  
Other aspects include: 
\begin{itemize}

\item The notion of change to the reference system contained already in earlier work by Galileo, Huygens, and Newton.
\item Rejection of Newton’s concept of absolute space, and hence adoption of the fully reciprocal relations (\ref{eqn:2}) and (\ref{eqn:3}) instead of the ‘one-way’ relation for absolute space in (\ref{eqn:31}).
\item Use of these simple, intuitive relations to inform and guide us to the Lorentz-type transformations.   Especially, through the method of changing a postulate that was introduced by Saccheri in 1733.
\item A study of these transformations to infer the existence of a universal constant $V^2$.
\item Appeal to an experiment, such as Fizeau’s experiment (1851), to infer that $V^2 = c^2$, and hence the Lorentz transformations.
\end{itemize}
The last four steps were, however, not carried out during these two centuries.  This despite Huygen’s conviction (1694) that choice of a reference system is essentially a matter of mathematical convenience.  However, this clue was overlooked by his successors, who failed to consider our key relations (\ref{eqn:2}) and (\ref{eqn:3}).  

Even though this was a golden age in Europe, well-endowed with great mathematicians interested also in ‘natural philosophy’ (among them the French savants, the Bernoullis, Euler, Gauss, and later, Riemann and Clifford).  People willing and able to go beyond ‘na\"ive visualizability’, to explore non-Euclidian geometry and its applications in Physics, encroaching even on aspects of the subsequent general relativity.  

In a prescient remark, Descartes stated that there, "is... no corresponding necessity that the external world be in anyway similar to the one our senses depict." [\onlinecite{75}].  Westfall rated this as "the most important assertion in natural philosophy during the entire seventeenth century." [\onlinecite{75}].  It seems fitting that Descartes' cautionary note (unheeded at the time) manifested first in a theory based on Cartesian axes and 'the modern' mathematics (analysis) that he pioneered.

Namely the relativity of time intervals and of simultaneity of spatially separated events [\onlinecite{6}]. For instance: "We believe we \textbf{know} what simultaneity means. And that is why the public response to the special theory of relativity was, and is, one of incredulity among both amateurs and professionals." [\onlinecite{76}].  According to Miller: "Although a change in the spatial dimensions of a body was acceptable to everyone, a relativity of time was not." [\onlinecite{77}].  Therefore, a portal to the special theory via relative space (Section II) would presumably be more evident than one proceeding via relative time.

\begin{acknowledgments}
In memory of my colleague, Professor Roger Raab (1934 --- 2022).  Thanks to Natasha Perry, Amanda Cocorozis, and Dr. Sarah Bryan for their essential help in preparing the manuscript.
\end{acknowledgments}

\appendix
\section{Transformation of Force}

The law of motion for a particle of (invariant) mass $m$ in an inertial frame $K$ is
\begin{equation}
\mathbf{F} = \frac{d\mathbf{p}}{dt} \, . \label{eqn:A.1}
\end{equation}

Here
\begin{equation}
\mathbf{p} = \gamma(u) \, m \, \mathbf{u} \, , \label{eqn:A.2}
\end{equation}
with $\gamma(u) = (1-u^2/V^2)^{-1/2}$, is the momentum of a particle moving with velocity
\begin{equation}
\mathbf{u} = \frac{d\mathbf{r}}{dt} \label{eqn:A.3}
\end{equation}
with respect to $K$.

In a second inertial frame, $K'$, the law of motion is
\begin{equation}
\mathbf{F}' = \frac{d\mathbf{p}'}{dt'} \, . \label{eqn:A.4}
\end{equation}
We seek the relations between the components of $\mathbf{F}$ and $\mathbf{F}'$ when the space and time coordinates are connected by the Lorentz-type transformations (\ref{eqn:12}), (\ref{eqn:14}), and (\ref{eqn:47}). Therefore, we require the transformations of $d/dt$ and $\mathbf{p}$.

Now
\begin{equation}
\begin{split}
\frac{d}{dt'} &= \frac{dt}{dt'} \frac{d}{dt} \\
&= \left\{\gamma(v)(1-vu_x/V^2)\right\}^{-1} \frac{d}{dt} \, ,
\end{split} \label{eqn:A.5}
\end{equation}
where we have used (\ref{eqn:14}).

In the momentum $\mathbf{p} = \gamma(u) \, m \, d\mathbf{r}/dt$, the ratio $dt/\gamma(u)$ is a standard invariant (the proper time interval). Thus $\mathbf{p}$ transforms in the same way as $d\mathbf{r}$:
\begin{equation}
\begin{split}
p_x' &= \gamma(u) \, m \, \frac{\gamma(v)(dx - v \, dt)}{dt} \\
&= \gamma(v)(p_x - v\gamma(u)m) \, ,
\end{split} \label{eqn:A.6}
\end{equation}
\begin{equation}
p_y' = p_y \, , \quad p_z' = p_z \, , \label{eqn:A.7}
\end{equation}
where we have used (\ref{eqn:14}) and (\ref{eqn:47}).

By substituting (\ref{eqn:A.5})--(\ref{eqn:A.7}) in (\ref{eqn:A.4}) we obtain
\begin{equation}
\mathbf{F}' = \frac{1}{1-vu_x/V^2} \left(\frac{dp_x}{dt} - v\frac{d}{dt}\gamma(u)m, \, \frac{1}{\gamma(v)}\frac{dp_y}{dt}, \, \frac{1}{\gamma(v)}\frac{dp_z}{dt}\right) \, . \label{eqn:A.8}
\end{equation}

A short calculation shows that for the force $\mathbf{F} = d\{\gamma(u)m\mathbf{u}\}/dt$,
\begin{equation}
\frac{d}{dt}\gamma(u)mV^2 = \mathbf{u} \cdot \mathbf{F} \, . \label{eqn:A.9}
\end{equation}
(That is, the rate of change of the particle's energy is equal to the rate at which $\mathbf{F}$ does work on the particle.) From (\ref{eqn:A.8}) and (\ref{eqn:A.9}) we obtain the desired force transformations:
\begin{equation}
F_x' = \frac{F_x - (v/V^2)\mathbf{u}\cdot\mathbf{F}}{1-vu_x/V^2} \, , \label{eqn:A.10}
\end{equation}
\begin{equation}
F_{y,z}' = \frac{F_{y,z}}{\gamma(v)(1-vu_x/V^2)} \, . \label{eqn:A.11}
\end{equation}

There are two useful special cases here. First, if the particle is instantaneously at rest in $K$ (i.e.\ $\mathbf{u}=0$), then (\ref{eqn:A.10}) and (\ref{eqn:A.11}) give
\begin{equation}
\mathbf{F} = (F_x', \, \gamma F_y', \, \gamma F_z') \, . \label{eqn:A.12}
\end{equation}
This relation is used to obtain the electric field $\mathbf{E}$ of a uniformly moving point charge; see (\ref{eqn:53}). Second, if $\mathbf{u}$ is parallel to $\mathbf{v}$ (i.e.\ $\mathbf{u} = u_x\hat{\mathbf{x}}$) then
\begin{equation}
\mathbf{F} = (F_x', \, \gamma(1-u_xv/V^2)F_y', \, \gamma(1-u_xv/V^2)F_z') \label{eqn:A.13}
\end{equation}
This relation is used in the derivation of the electric and magnetic forces between charges moving along parallel lines, see (\ref{eqn:57}).

If the direction of $\mathbf{u}$ is arbitrary, then the contributions of $u_yF_y$ and $u_zF_z$ in (\ref{eqn:A.10}) must be included. The $y$- and $z$-components of (\ref{eqn:A.13}) are unchanged (see (\ref{eqn:A.11})), but the $x$-component is replaced by
\begin{equation}
F_x' + (\gamma u_y v/V^2)F_y' + (\gamma u_z v/V^2)F_z' \, . \label{eqn:A.14}
\end{equation}
The terms in $u_y$ and $u_z$ produce the $x$-component of (\ref{eqn:58}).
\nocite{*}
\bibliography{ref}
\end{document}